# Crystal and magnetic structure of antiferromagnetic Mn$_2$PtPd

Vivek Kumar[1], Manfred Reehuis[2], Andreas Hoser[2], Peter Adler[1], and Claudia Felser[1]

[1]*Max Planck Institute for Chemical Physics of Solids, Nöthnitzer Str. 40, D-01187 Dresden, Germany*

[2]*Helmholtz-Zentrum Berlin für Materialien und Energie, D-14109 Berlin, Germany*

**Abstract**

We have investigated the crystal and magnetic structure of Mn$_2$PtPd alloy using powder X-ray and neutron diffraction experiments. This compound is believed to belong to the Heusler family having crystal symmetry *I*4/*mmm* (TiAl$_3$-type). However, in this work we found that the Pd and Pt atoms are disordered and thus Mn$_2$PtPd crystallizes in the *L*1$_0$ structure having *P*4/*mmm* symmetry (CuAu-I type) like MnPt and MnPd binary alloys. The lattice constants are $a$ = 2.86 Å and $c$ = 3.62 Å at room temperature. Mn$_2$PtPd has a collinear antiferromagnetic spin structure below the Néel temperature $T_N$ = 866 K, where Mn moments of ~4 $\mu_B$ lie in the *ab*-plane. We observed a strong change in the lattice parameters near $T_N$. The sample exhibits metallic behaviour, where electrical resistivity and carrier concentration are of the order of 10$^{-5}$ Ω cm and 10$^{21}$ cm$^{-3}$, respectively.

**1. Introduction**

Antiferromagnets (AFMs) are special among magnetic materials as they display magnetic ordering but with zero magnetic moment. Their practical use has been established in many fields especially in spintronics, for instance as pinning layers in giant magnetoresistance (GMR) and tunnel magnetoresistance (TMR) devices [1–13]. Here, the AFM acts as a passive component. The absence of stray fields is a great advantage of AFMs over ferromagnets and thus AFMs may even replace ferromagnets as an active component in spintronic devices. However, it is difficult to manipulate the AFMs due to their vanishing magnetic moment. Recent advances in controlling the antiferromagnetic configurations by electrical switching bear a good prospect for a new era of applications [14]. In addition, high ordering temperatures and large magneto-crystalline anisotropy are desired to enhance the scope for more universal use.

Heusler alloys *X*$_2$*YZ* are a particularly promising class of materials for the search of new AFMs for applications due to their great variability in chemical composition and properties [15–20]. In this regard the recently reported phase Mn$_2$PtPd is of interest, which was identified in a high throughput computational study by Sanvito *et al.* [21] as a potential new ferromagnetic cubic

Heusler alloy. Its subsequent experimental realization, however, rather suggested this material to be a tetragonal antiferromagnetic Heusler compound having space group $I4/mmm$ (TiAl$_3$-type) with lattice constants $a = 4.03$ Å and $c = 7.24$ Å [21]. A TiAl$_3$-type structure would imply ordering of Pt and Pd atoms. A magnetic transition near 320 K was reported [21], but details on the magnetic behaviour are unknown yet. On the other hand, several Mn based binary alloys Mn$TM$ with various transition metals $TM$ like Ni, Pt, Pd, Rh, Ir have been reported which are AFMs having high Néel temperatures and the crystal symmetry $P4/mmm$ ($L1_0$, CuAu-I type) [22–32]. Binary alloys having cubic ($B2$, CsCl-type) and tetragonal ($L1_0$, CuAu-I type) structures are the building blocks to design Heusler materials [33]. Combining two binary alloys $XY$ and $XZ$ belonging to the $B2$ structure results in a regular Heusler ($L2_1$, Cu$_2$MnAl-type) structure, given the condition that $Y$ and $Z$ atoms order, whereas a tetragonal Heusler (TiAl$_3$-type) compound is expected in the case of combining two $L1_0$ structures. However, ternary alloys $X_2YZ$ adapt the same structure as their binary precursor if there is a complete disorder of $Y$ and $Z$ atoms. Hence, taking into account the chemical similarity between Pd and Pt atoms, Mn$_2$PtPd could also be considered as a CuAu-I type phase MnPt$_{0.5}$Pd$_{0.5}$ with a disordered arrangement of Pt and Pd atoms. In this paper we report our powder X-ray and neutron diffraction studies on Mn$_2$PtPd as well as magnetization and electrical transport measurements. We will show that Mn$_2$PtPd is a CuAu-I type antiferromagnet with a Néel temperature $T_N = 866(5)$ K.

## 2. Experimental details

A polycrystalline ingot of Mn$_2$PtPd was prepared by arc melting stoichiometric amounts of constituent elements in the presence of high purity Ar atmosphere. About 2.5 wt % of extra Mn was used to compensate the weight loss due to the evaporation of Mn during the melting. The sample was ground and characterized at room temperature by powder X-ray diffraction (XRD) using a Huber G670 camera [Guinier technique, λ = 1.54056 Å (Cu−K$α_1$ radiation)]. Field and temperature dependent magnetization measurements were performed using a vibrating sample magnetometer (MPMS3, Quantum Design). Temperature dependent magnetization $M(T)$ measurements were carried out from 2 to 400 K in zero field cooled (ZFC) and field cooled (FC) modes. High temperature magnetization measurements were performed during heating and cooling from 300 to 1000 K using the oven option of the MPMS3. The electrical transport properties were investigated using a physical property measurement system (PPMS, Quantum Design). The Hall measurements were performed on a rectangular bar using a five probe geometry at different temperatures from 2 K to 300 K in fields up to 5 T. For the neutron diffraction study, the sample was powdered by grinding followed by annealing at 773 K in an evacuated quartz tube for 24 hours. Neutron powder diffraction experiments were carried out on the instrument E6 at the BER II reactor of the Helmholtz-Zentrum Berlin. This instrument uses a pyrolytic graphite (PG) monochromator selecting the neutron wavelength $\lambda = 2.42$ Å. In order to investigate in detail the crystal structure of $Mn_2PtPd$ a powder pattern was collected at 1012 K, well above the magnetic ordering temperature of $T_N = 866$ K. The sample was heated up in a quartz ampoule using a high-temperature furnace (AS Scientific Products Ltd., Abingdon, GB). The temperature dependence of the crystal and magnetic structure was investigated between 296 and 1012 K, where 18 and 6 patterns were collected below and above $T_N$, respectively. Neutron powder patterns were recorded between the diffraction angles ($2\theta$) 5.5 and 136.5 °. Rietveld refinements of the powder diffraction data were carried out with the program *FullProf* [34], using the nuclear scattering lengths $b(Mn) = -3.73$ fm, $b(Pd) = 5.91$ fm, and $b(Pt) = 9.63$ fm [35]. The magnetic form factor of the Mn atoms was taken from Ref. [36].

## 3. Results and discussion

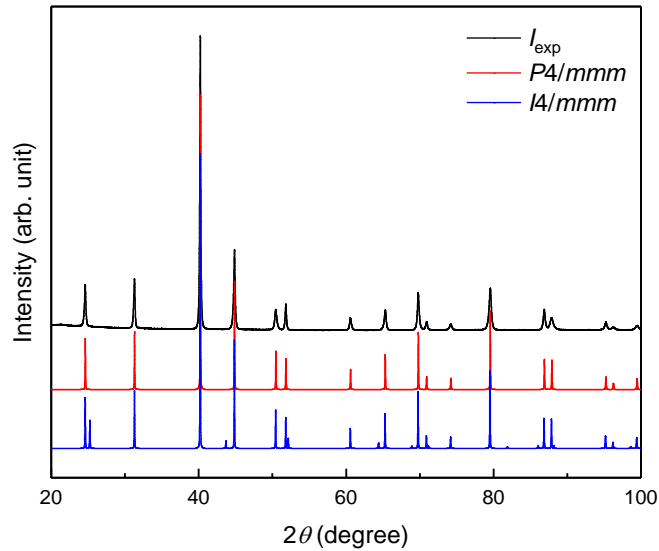

**Figure 1.** Observed X-ray diffraction pattern (black) of $Mn_2PtPd$. For comparison, calculated patterns for space groups *P*4/*mmm* (CuAu-I type, red) and *I*4/*mmm* (TiAl$_3$-type, blue) are shown together.

The phase purity of the sample was examined by powder X-ray diffraction as shown in figure 1. To determine the crystal structure, calculated XRD patterns for both structure models *P*4/*mmm* (CuAu-I type with $a = 2.86$ Å and $c = 3.61$ Å) and *I*4/*mmm* (TiAl$_3$-type with $a = 4.04$ Å and $c = 7.23$ Å) are shown in addition. The observed diffraction peaks match well with both models. In particular, the lattice parameters compare well with those reported by Sanvito *et* *al.* [21] using space group *I*4/*mmm* (TiAl$_3$-type, *a* = 4.03 Å and *c* = 7.24 Å). However, the absence of some additional peaks in the experimental pattern which are present in the calculated *I*4/*mmm* pattern suggests that Pt and Pd are atomically disordered and thus Mn$_2$PtPd rather crystallizes in *P*4/*mmm* symmetry. Below we will show the detailed structural analysis using the powder neutron diffraction data.

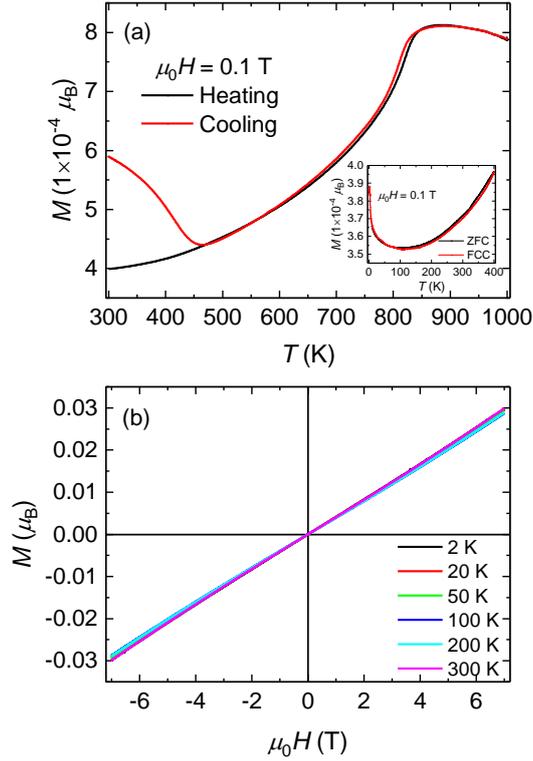

**Figure 2.** (a) Temperature dependent magnetization *M*(*T*) per formula unit of Mn$_2$PtPd from 300 to 1000 K measured using the oven option and from 2 to 400 K (in the inset) (b) Magnetic isotherms at different temperatures.

To determine the magnetic behaviour and the ordering temperature we have performed magnetization measurements as a function of temperature and field. Figure 2(a) shows the temperature dependence of the magnetization *M*(*T*) from 300 to 1000 K and in the inset from 2 to 400 K. For the high temperature *M*(*T*) measurements the sample was heated from room temperature to 1000 K and then cooled down to room temperature again in the presence of a magnetic field of 0.1 T. The broad maximum in the *M*(*T*) curve indicates that the sample undergoes an antiferromagnetic ordering transition with a Néel temperature (*T*$_N$) of about 870 K. The observed *T*$_N$ for Mn$_2$PtPd is close to the average of the *T*$_N$ values of 780 and 970 K which were reported for the binary AFMs MnPd and MnPt, respectively [25,26,29–31]. This is further support that Mn$_2$PtPd rather is a CuAu-I type than a TiAl$_3$-type AFM material. For obtaining low temperature *M*(*T*) data, the sample was cooled down to 2 K and measurements were carried out in the zero field cooled (ZFC) and field cooled (FC) modes from 2 to 400 K at 0.1 T. Both ZFC and FC magnetization curves followed the same path. In contrast to Ref. [21] we did not observe any indication for a magnetic transition near 320 K. At temperatures below 100 K an up-turn in the magnetization is apparent. A similar behaviour has been reported for MnPt alloys due to a very small deviation from the equiatomic composition [26]. Therefore, a small amount of off-stoichiometry in the sample could be the reason for this kind of behaviour and also for the irreversibility in the magnetization being observed at 470 K on cooling (figure 2(a)). In figure 2(b) magnetic isotherms *M*(*H*) are shown at different temperatures. The linear increase of *M* with *H* is in agreement with the anticipated AFM order. There are no indications for a hysteresis or spontaneous magnetization.

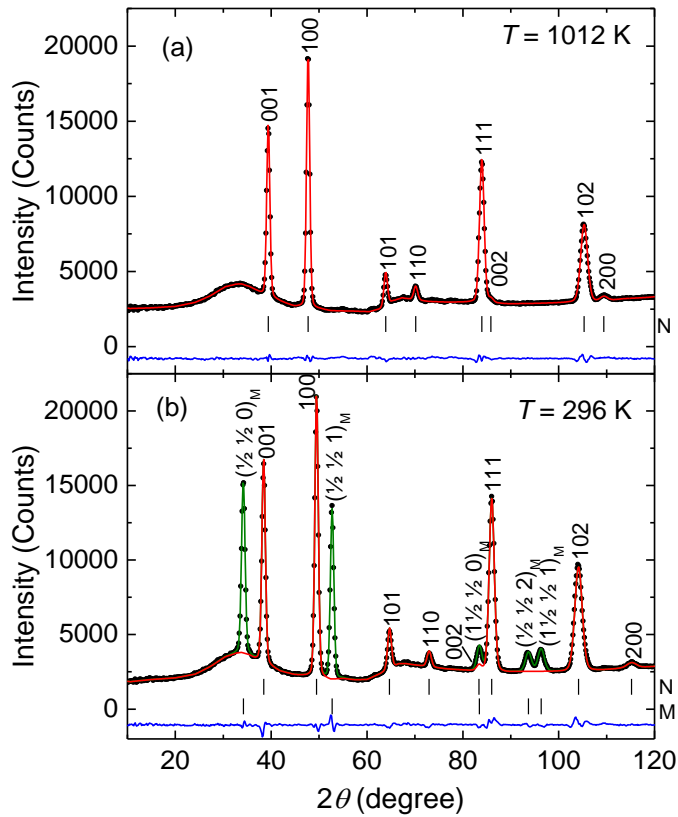

**Figure 3.** Results of the Rietveld refinements of the neutron powder diffraction data of $Mn_2PtPd$ collected on the instrument E6 at 1012 K (a) and 296 K (b), respectively. The crystal structure was refined in the tetragonal space group *P*4/*mmm*. The calculated patterns (crystal structure in red, crystal and magnetic structure in green) are compared with the observed ones (black circles). The difference patterns (blue) as well as the positions (black bars) of the nuclear (N) and magnetic (M) Bragg reflections are shown. The broad features are due to scattering by the quartz sample container.

The crystal structure of Mn$_2$PtPd was refined from a powder neutron diffraction data set collected at 1012 K (figure 3(a)), which is above the magnetic ordering temperature of $T_N$ = 866 K. The refinements were carried out in the tetragonal space groups *P*4/*mmm* (No. 123) and *I*4/*mmm* (No. 139), respectively. In the space group *P*4/*mmm* the Mn atoms are located at the Wyckoff position 1*d*(½,½,½), while the Pd and Pt atoms are statistically distributed at the position 1*a*(0,0,0). The refinement resulted in a satisfactory residual $R_F$ = 0.0178. In the space group *I*4/*mmm* the unit cell is enlarged with the dimensions $a\sqrt{2} \times a\sqrt{2} \times 2c$. In this setting the Mn atoms occupy the position 4*d*(0,½,¼), while the Pd and Pt occupy different atomic sites located at the positions 2*a*(0,0,0) and 2*b*(0,0,½) or vice versa. The refinement resulted exactly in the same residual for both settings, but the obtained residual of $R_F$ = 0.0357 is considerably enlarged compared to the refinement in *P*4/*mmm*. This can be ascribed to the fact that a segregation of the Pt and Pd atoms clearly leads to an emergence of the Bragg reflections 101, 103, and 213 (½ ½ 1½, ½ ½ 1½, and 1½ ½ 1½ in the setting of *P*4/*mmm*). Due to the fact that these reflections do not exist the Pd and Pt atoms are statistically distributed at both positions 2*a* and 2*b*. Considering the fact that no intensity was detected at these particular reflections we conclude that the crystal structure of Mn$_2$PtPd can be well described in the space group *P*4/*mmm* having the smaller unit cell. The results of the Rietveld refinements are summarized in table 1.

**Table 1.** Results of the neutron diffraction study of Mn$_2$PtPd. The crystal structure refinements of the data sets were carried out in the tetragonal space group *P*4/*mmm*. Listed are the unit cell parameters and the interatomic distances between the metal atoms (up to 4 Å) obtained at 296 and 1012 K. Further the experimental magnetic moment of the Mn atom is given. The listed residuals of the refinement of the crystal and magnetic structure are defined as $R_F = \sum ||F_{obs}| - |F_{calc}||/\sum|F_{obs}|$ and $R_M = \sum ||I_{obs}| - |I_{calc}||/\sum|I_{obs}|$, respectively.

| Mn$_2$PtPd | at 296 K | at 1012 K |
| --- | --- | --- |
| *a* [Å] | 2.85512(13) | 2.95250(13) |
| *c* [Å] | 3.61438(18) | 3.53069(18) |
| *c*/*a* | 1.26593(11) | 1.19583(11) |
| *V* [Å$^3$] | 29.463(3) | 30.778(3) |
| *d*(Mn-Pt/Pd) | 2.7096(1) | 2.7341(1) |
| $d_e$(Mn-Mn) | 2.8551(1) | 2.9525(1) |
| $d_a$(Mn-Mn) | 3.6144(2) | 3.5307(2) |
| $R_F$ | 0.0178 | 0.0210 |
| $\mu_{exp}$(Mn) | 4.093(10) | - |
| $R_M$ | 0.0214 | - |

In the neutron powder patterns collected at 296 K (figure 3(b)) two strong magnetic reflections were observed at the $2\theta$ positions 34.8 and 53.4°. This can be ascribed to a long-range magnetic ordering of the Mn moments. These reflections could be indexed as (½ ½ 0)$_M$ and (½ ½ 1)$_M$ using the smaller unit cell (crystal structure type with space group P4/*mmm*) as described above. These two reflections can be generated by the rule $(hkl)_M = (hkl)_N \pm k$, where the propagation vector is $k = (½,½,0)$. This suggests that the magnetic unit cell parameters are doubled along the *a* and *b* axes. In this type of magnetic ordering the magnetic moments of the Mn atoms are antiferromagnetically coupled along the *a* and *b* axes with the spin sequence + − + − ... . From a symmetry analysis using the program BASIREP of the FullProf suite [34] one obtains two irreducible representations, which describe a spin structure, where the moments are either aligned parallel to the tetragonal *c* axis or aligned within the *ab* plane, respectively. Our Rietveld refinements clearly show that the Mn moments are aligned within the tetragonal *ab* plane, resulting in a residual $R_M = 0.0214$, where the intensity ratio of the first two magnetic reflections is $I(½ ½ 0)_M/I(½ ½ 1)_M = 1.0$. Assuming a spin alignment parallel to the *c* axis one obtains a considerably enlarged ratio of 3.3. In our refinements we further assumed the magnetic moments to be aligned parallel to the directions [110] or [100]. For these two models we practically obtained the same residuals, which shows that we are not able to determine the moment direction within the tetragonal *ab* plane. It has been reported that MnPt exhibits two different magnetic structures [30]. At high temperatures above 750 K, the Mn moment lies in the basal plane whereas it undergoes a spin-flip transition in a broad temperature region ranging from 750 to 570 K, leading to a magnetic structure with Mn moments aligned parallel to the tetragonal *c* axis. By contrast, MnPd exhibits only the magnetic structure with the moments lying in the *ab* plane.

The same spin structure is also adopted by Mn$_2$PtPd throughout the whole temperature range. There are no indications for a spin-flip transition. It is noted that we did not observe any significant change in the magnetic configuration below 470 K which indicates that the irreversibility in the *M*(*T*) curve (figure 2a) is rather related to off-stoichiometry or defects than being an intrinsic bulk property.

The determined moment value at 296 K is $\mu_{exp}$(Mn) = 4.09(10) $\mu_B$, which is similar as for many other magnetic intermetallic alloys containing Mn [24–26,29,31,37]. The temperature dependence of the magnetic moment of the Mn atom was investigated in the temperature range from 296 to 1012 K. In figure 4(a) it is seen that the saturation of the magnetic moments is almost reached at 296 K. The magnetic order disappears at the Néel temperature $T_N$ = 866(5) K. The lattice parameters, the $c/a$ ratio, and the volume of the unit cell are displayed in figure 4 (b) and 4 (c) respectively. The lattice parameters at 296 K are $a$ = 2.86 Å and $c$ = 3.62 Å. Interestingly, the Rietveld refinements of the powder patterns revealed a strong change of the lattice parameters near the Néel temperature, which results in a pronounced increase of the tetragonal distortion as reflected in the $c/a$ ratio. This is ascribed to the fact that the apical bond length $d_a$(Mn-Mn) is strongly elongated below $T_N$, whereas the equatorial bond length $d_e$(Mn-Mn) is shortened (see table 1). A similar strong magnetoelastic coupling on going from the paramagnetic to the AFM phase was observed for MnPd and the lattice anomaly was attributed to deformation sensitive nearest neighbour Mn-Mn interactions [29]. On the other hand the cell volume is continuously decreasing to lower temperatures without pronounced anomalies near the Néel temperature.

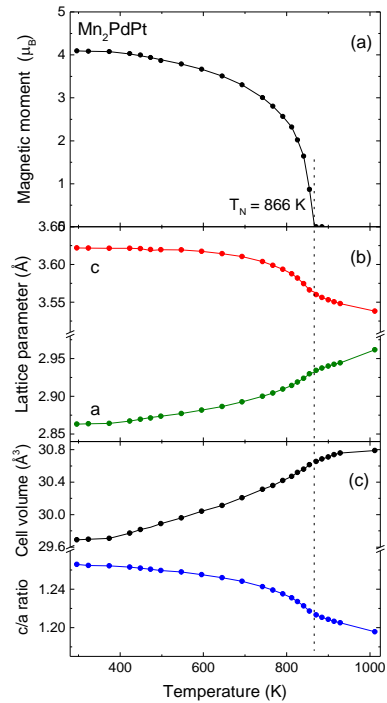

**Figure 4.** (a) Temperature dependence of the magnetic moment of the Mn atoms in Mn$_2$PtPd. Antiferromagnetic ordering sets in at the Néel temperature $T_N$ = 866(5) K, where the lattice parameters $a$ and $c$ [shown in (b)] reveal a strong decrease and increase, respectively. The temperature dependence of the $c/a$ ratio and the unit cell volume $V$ are shown in (c). The dotted lines are only a guide for the eye.

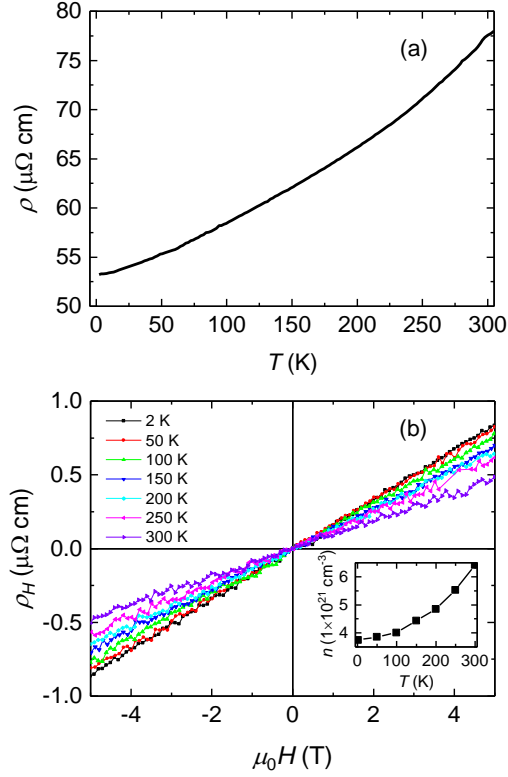

**Figure 5.** (a) Temperature dependence of electrical resistivity $\rho(T)$ and (b) Hall resistivity $\rho_H$ as a function of field from 2 to 300 K of Mn$_2$PtPd. The carrier concentration $n$ is shown in the inset.

Figure 5 (a) shows the electrical resistivity as a function of temperature from 2 K to room temperature. The sample exhibits metallic behavior and has resistivity values of the same order of magnitude ($10^{-5}$ $\Omega$ cm) as those of MnPt [26] and MnPd alloys [25]. We measured the Hall resistivity $\rho_H$ at different temperatures between 2 and 300 K as shown in figure 5 (b). The straight line behaviour of the Hall resistivity with the applied field is expected for the collinear AFM structure which shows a normal Hall effect. Anomalous Hall effects scaling with the magnetization are only found for ferromagnets or peculiar types of noncollinear antiferromagnets [38]. The calculated carrier concentration from Hall measurements is shown in the inset of figure 5 (b). The carrier concentration is given by the equation $n = \mu_0 H/e\rho_H$, where $\mu_0$ is the vacuum permeability, $H$ is the applied field and $e$ is the elementary charge. The carrier concentration is of the order of $10^{21}$ cm$^{-3}$. The Hall resistivity decreases and carrier concentration increases with increase in temperature.

## 4. Conclusions

We found that Mn$_2$PtPd adopts the $L1_0$ (CuAu-I)-type crystal structure similar to MnPt and MnPd binary alloys. As for MnPd the magnetic moments of the Mn atoms (~4 $\mu_B$) are antiferromagnetically coupled in the *ab*-plane without evidence for a spin-flip transition. Mn$_2$PtPd has an ordering temperature $T_N$ = 866 K which is the average of the ordering temperatures of MnPd and MnPt. A pronounced change in the lattice parameters near $T_N$ shows strong magnetoelastic coupling. We observe that Mn$_2$PtPd has metallic characteristics analogous to MnPt and MnPd.

**Acknowledgment**

This work was financially supported by the ERC Advanced Grant "TOPMAT" (No. 742068). We thank Walter Schnelle and Ralf Koban for the high temperature magnetization measurements and Horst Borrmann for XRD measurements.